# Low-Frequency Noise and Sliding of the Charge Density Waves in Two-Dimensional Materials

Guanxiong Liu[1], Sergey Rumyantsev[1,2], Matthew. A. Bloodgood[3], Tina T. Salguero[3] and Alexander A. Balandin[1*]

[1]Nano-Device Laboratory (NDL) and Phonon Optimized Engineered Materials (POEM) Center, Department of Electrical and Computer Engineering, Materials Science and Engineering Program, University of California, Riverside, California 92521 USA

[2]Ioffe Physical-Technical Institute, St. Petersburg 194021 Russia

[3]Department of Chemistry, University of Georgia, Athens, Georgia 30602 USA

There has been a recent renewal of interest in charge-density-wave (CDW) phenomena, primarily driven by the emergence of two-dimensional (2D) layered CDW materials, such as 1$T$-TaS$_2$, characterized by very high transition temperatures to CDW phases[1–16]. In the extensively studied *classical* bulk CDW materials with quasi-1D crystal structure, the charge carrier transport exhibits intriguing *sliding* behavior, which reveals itself in the frequency domain as "narrowband" and "broadband" noise[17–22]. Despite the increasing attention on physics of 2D CDWs, there have been few reports of CDW sliding, specifically in quasi-2D rare-earth tritellurides[2] and *none* on the noise in any of 2D CDW systems. Here we report the results of low-frequency noise (LFN) measurements on 1$T$-TaS$_2$ thin films - archetypal 2D CDW systems, as they are driven from the nearly commensurate (NC) to incommensurate (IC) CDW phases by voltage and temperature stimuli. We have found that noise in 1$T$-TaS$_2$ devices has *two* pronounced maxima at the bias voltages, which correspond to the onset of CDW sliding and the NC-to-IC phase transition. We observed unusual Lorentzian noise features and *exceptionally* strong noise dependence on electric bias and temperature. We argue that LFN in 2D CDW systems has unique physical origin, different from known fundamental noise types. The specifics of LFN in 2D CDW materials can be explained by invoking the concept of interacting discrete fluctuators in the NC-CDW phase. Noise spectroscopy can serve as a useful tool for understanding electronic transport phenomena in 2D CDW materials characterized by coexistence of different phases and strong CDW pinning.

---

[*] Corresponding author (AAB): balandin@ece.ucr.edu ; http://balandingroup.ucr.edu/

## Main Text

The CDW phase is a macroscopic quantum state consisting of a periodic modulation of the electronic charge density accompanied by a periodic distortion of the atomic lattice[17]. The early work on CDW effects was performed with *bulk* samples, e.g. $TaS_3$ or $NbSe_3$ crystals, which have quasi-one-dimensional (1D) crystal structures of strongly-bound 1D atomic chains that are weakly bound together by van-der-Waals forces[17–22]. One of the most interesting phenomena observed in bulk quasi-1D CDW materials is the sliding of CDWs when the applied electric field exceeds a threshold value, $E_{DT}$, sufficient to de-pin the CDWs from defects[17]. CDW sliding leads to nonlinear DC conductivity, "narrowband noise", which has the frequency, $f$, proportional to the current, $I$, and "broadband noise"[18–22].

In recent years there has been a rebirth of interest in physics of CDW effects, with the focus on different type of materials, such as layered 2D crystals of $1T$-$TaS_2$ and $1T$-$TaSe_2$, which are members of the transition metal dichalcogenide (TMD) family. Unlike classical bulk CDW materials, these TMDs exhibit unusually high transition temperatures to different CDW symmetry-reducing phases[1,5,10–15]. The most interesting of these materials, $1T$-$TaS_2$, undergoes a transition from the normal metallic phase to the IC-CDW phase at 545 K, then to the NC-CDW phase at 355 K, and finally to the commensurate (C) CDW phase at 180 K[11–15]. The NC-CDW phase consists of commensurate domains separated by regions that are incommensurate with the underlying crystal lattice. The transitions among CDW phases in $1T$-TaS$_2$ films can be affected by the number of layers as well as the applied field[11,14]. The de-pinning and sliding of CDWs in quasi-2D materials is much less explored than in bulk quasi-1D CDW materials. Earlier attempts in finding the sliding effect in quasi-2D materials were not successful[23]. We are aware of just few reports on CDW sliding in only one specific material system - quai-2D rare-earth tritellurides[2] – and none on either "broadband" or "narrowband" noise in any of 2D CDW systems, despite the importance of such characteristics for understanding CDW effects.

In this Letter, we report on the low-frequency current fluctuations, *i.e.* LFN, in an archetypal 2D CDW material system, $1T$-TaS$_2$, considering both frequency and time-domain signals. This type of fluctuation, typically with the spectral density $S(f) \sim 1/f^{\gamma}$ ($\gamma \approx 1$), is found in almost all materials and devices[24]. It is conventionally known as $1/f$-noise in electronics, and it often is referred to as "broadband" noise in the CDW field to distinguish it from "narrowband" noise[17–22]. Although practical applications can benefit from the reduction of LFN, it also can be used as an important metric that reveals information about the physical processes in materials. Here we use LFN measurements to unambiguously observe the de-pinning and sliding of CDWs in the NC-CDW phase of $1T$-TaS$_2$. Our results reveal *unusual* features in the LFN spectra of this quasi-2D CDW system, which are strikingly different from those in metals or semiconductors.

The high-quality $1T$-TaS$_2$ two-terminal devices were fabricated from thin films, exfoliated from single crystals grown by the chemical vapor transport method (see *Methods*). The thickness of the thin films was ~20 nm. The films were capped with $h$-BN to protect the $1T$-TaS$_2$ channels from degradation[14]. The contacts were defined by electron beam lithography, resulting in devices with lateral dimensions of 1-3 μm by 0.8-1.5 μm. The *edge* contacts were fabricated by depositing metal after plasma etching through the $h$-BN capping layer. Fig. 1a shows DC $I$-$V$ characteristics of a representative $1T$-TaS$_2$ device at room temperature (RT). The CDW phase transition from NC to IC states is observed as an abrupt jump in the current and accompanying hysteresis, when the voltage exceeds the threshold voltages $V_H$ and $V_L$. Whereas the NC-IC CDW transition is visible in the *I-V* curve, the de-pinning and onset of CDWs sliding is not. Only from theoretical considerations can one deduce that somewhere before the hysteresis, where *I-V* becomes *superliner* (*e.g.* regions I, II or III), the collective current of the sliding CDW starts to contribute to the total current[17]. Even the derivative, *dI/dV*, does not identify the exact onset of sliding (*Supplemental Materials*). This feature is in striking contrast to the behavior of bulk quasi-1D CDW materials, which are characterized by an abrupt transition from the single-particle to the collective current[17]. This difference can be attributed to the fact that the quasi-2D NC-CDW phase consists of a mixture of C-CDW islands surrounded by the more strongly conducting IC-CDW phase[11–14]. As a result, the onset of sliding is overshadowed by conduction via the IC-CDW continuum.

[Figure 1]

The LFN measurements were conducted under DC bias voltages, $V_b$, increasing from 30 mV to 1.09 V, beyond the phase transition voltage ($V_H = 0.92$ V for this device). The DC bias points for each noise measurement are indicated by the red circles in Fig. 1a. The normalized current-noise-spectral-density, $S_I/I^2$, *vs.* $V_b$ (*f*=10 Hz), measured at RT, is presented in Fig. 1b. One can see *two* well-resolved local maxima, which were reproducible for all studied devices. The first maximum at $V_{DT}$=0.2V indicates the de-pinning and onset of CDW sliding, while the second marks the NC-IC phase transition. The voltage, $V_{DT}$, corresponds to the threshold field of $E_{DT}$=1.3 kV/cm for this device, a value that is four orders-of-magnitude larger than that in typical bulk quasi-1D CDW materials[17]. Considering that $E_{DT} \sim A^{-1/2}$ (*A* is the cross-section area)[17], we conclude that the large threshold field in quasi-2D materials is likely related to CDW pinning to the top and bottom surfaces. Fig. 1b indicates that LFN is an *unambiguous* metric for CDW sliding in 2D CDWs.

Fig. 1c-1f show $S_I/I^2$ *vs.* *f* for four different biasing regions, marked as I, II, III, and IV in Fig. 1a. At the small bias, $V_b \leq 150$ mV, $S_I/I^2$ follows typical $1/f$ dependence ($\gamma \approx 1$) as seen in Fig. 1c. When $V_b$ increases above 150 mV, the noise level increases sharply, and an excess noise with $\gamma = 2$ appears at the lower frequency range. As $V_b$ increases further, the noise spectrum evolves into a Lorentzian shape, $S_I(f)=S_o \times f_c^2/(f_c^2+f^2)$, where $f_c=(2\pi\tau)^{-1}$ is the corner frequency, $S_o$ is the frequency-independent portion of $S_I(f)$ observed at $f<f_c$, and $\tau$ is the characteristic time of the fluctuation process[25]. In a very narrow bias range from 130 mV to 180 mV, the LFN level increases by *four* orders-of-magnitude. This fast growth of LFN is a signature of CDW de-pinning[18–22]. Even more interestingly, $f_c$ shifts upwards with increasing bias, $V_b$ (see Fig. 1d). As $V_b$ approaches the NC-IC phase transition voltage $V_H$, the overall noise magnitude attains its second local maximum. This behavior is expected because the domains of the superstructure in the NC phase are experiencing a drastic lattice reconstruction and melting into the IC phase. Inside the hysteresis window (shown in Fig. 1a), $S_I(f)$ keeps almost the same shape with only small magnitude variations (Fig. 1e). When $V_b$ drives the transition to the IC-CDW phase, the noise decreases sharply, and the spectrum regains the $1/f$ dependence typical for metals (Fig. 1f). A sharp increase of noise near and at the phase transition, known for other materials[26,27], can be associated with abrupt changes in the resistance and instability of the phase transition. The strong dependence of noise on $V_b$ in the regions where there are small changes in *I-V* is more intriguing. Both the noise amplitude and spectrum shape change drastically with a very small change in $V_b$ in region II (Fig. 1b-f). The dependence of $f_c$ on $V_b$ for the region of the sliding CDW (region in II in Fig.1a) is summarized in Fig. 1g. A voltage increase of only 120 mV results in a *four* orders-of-magnitude change in $f_c$. This drastic change in $f_c$ with the bias is highly *unusual* for conventional materials, where a Lorentzian spectrum is associated with the generation-recombination (G-R) noise with $f_c$ independent from the bias[25]. Therefore, the nature of the Lorentzian-type spectra in Fig. 1 is different from the conventional G-R mechanism.

To elucidate the relative effects of electric field, *E*, and temperature, *T*, on CDW sliding in 2D systems, we performed noise measurement for *T* ranging from 290 K to 375 K covering the NC-IC phase transition. The experiments were conducted at two bias voltages $V_b$=30 mV and $V_b$=60 mV, with one less than than $V_{DT}$ and another greater than $V_{DT}$ ($V_{DT}$=50 mV) The temperature of the NC-IC phase transition, $T$=355 K, is the same for these two voltages (Fig. 2a). The coincidence of the phase-transition temperatures indicates that Joule heating is negligible at these small biases. The color map in Fig. 2b shows the evolution of $S_I/I^2 \times f$, with temperature and frequency at fixed $V_b$= 30 mV. Below 320 K, the noise is of $1/f$ – type and the magnitude of $S_I/I^2 \times f$ is small (blue color). At $T \approx$325 K, noise increases sharply, accompanied by a Lorentzian bulge with $f_c$=52 Hz (red color). As $T$ increases from 325 K to 375 K, $f_c$ remains almost constant (black arrow). Fig. 2c shows the temperature dependent noise profile at $f$=10 Hz. The noise level increases by four orders-of-magnitude at 325 K followed by a slow increase with increasing temperature towards the NC-IC phase transition. Upon entering the metallic-like IC phase, the noise decreases by one order-of-magnitude and continues to decrease with $T$. Temperature alone cannot change $f_c$ at the voltage below $V_{DT}$. In contrast, evolution of the Lorentzian bulge is completely different under the bias exceeding $V_{DT}$. Fig. 2d shows the color map of $S_I/I^2 \times f$ at $V_b$= 60 mV. The corresponding profile is presented in Fig. 2e. From ~840 Hz at 290 K to ~74 kHz at 315 K, $f_c$ increases quickly with temperature. At temperatures nearing the NC-IC phase transition, $f_c$ becomes undetectable. Above the NC-IC transition, the noise decreases and returns to $1/f$ – type. The evolution of $f_c$ with temperature at two biases is summarized in Fig. 2f. Only in the case of $V_b$>$V_{DT}$ does the Lorentzian corner frequency $f_c$ changes with temperature. We further verified that $f_c$ shifting was driven primarily by the electric field, by confirming that the Joule heating was negligible in these experiments (*Supplemental Materials*).

[Figure 2]

The temperature dependence of $f_c$ in semiconductors is often exponential, allowing for extraction of the activation energy, $E_a$, of traps responsible for G-R noise[25,28,29]. Using the data for $V_b$=60 mV in Fig. 2f, which correspond to the sliding-wave regime, one can extract $E_a \approx$2.3 eV. This energy is unrealistically high (larger than the bandgap in many semiconductors) to be compatible with a conventional G-R mechanism[25]. The bandgap of the C-CDW domains is about 0.2-0.4 eV[1], much smaller than the extracted $E_a$. Thus, LFN in this material system has a unique origin associated with the certain domains in NC-CDW phase and their evolution under electric and temperature stimuli. Systems that contain few fluctuators demonstrate random telegraph signal (RTS) noise. RTS noise with Lorentzian spectrum has been observed in many nanoscale devices [28,29], and it also has been encountered in bulk quasi-1D CDW systems[18,19]. We performed time-domain measurements for $V_b$ between 100 mV and 121 mV, where 1$T$-TaS$_2$ device started to show Lorentzian spectra (Fig. 3a). At $V_b$=100 mV, only "tails" with 1/$f^2$-dependence could be recorder because $f_c$ was below the measurement range. We found that at these biases, 1$T$-TaS$_2$ reveal large levels of the multi-state RTS noise. Figs. 3b-d show the RTS noise at different time scales as $V_b$ increases from 100 mV to 121 mV. The occurrence of RTS noise indicates that de-pinning takes place in just a few domains. Interestingly, a very small $V_b$ increase from 100 mV to 106 mV (Fig. 3b) leads to a significant change in the current fluctuations. The amplitude of the pulses increases and the number of levels becomes greater than three, meaning that the number of de-pinned domains increases sharply with increasing bias. This scenario is drastically different from classical RTS noise in semiconductor devices where biasing conditions usually change the pulses duration and intervals between pulses but not the amplitude or number of levels[28–30].

[Figure 3]

If we consider a shorter time scale, the traces for the same bias, $V_b$=106 mV, again look like two- or three-level RTS (Fig. 3c). This is because, at a short time scale, the rare high-amplitude events are not captured. A small increase in $V_b$ results in the same changes, however. The time traces for further-reduced timescales show a consistent trend (Fig. 3d). The bias voltage increases the number of RTS levels, $i.e.$ discrete fluctuators. At $V_b$=116 mV and 121 mV, the time traces in this timescale are not standard RTS because there are too many levels (Fig. 3d). However, the noise is still non-Gaussian because the time traces are highly asymmetric[28–30]. For $V_b$= 121 mV, the high amplitude peaks can only be attained after a series of pulses with smaller amplitudes. This suggests that $V_b$ changes the number of discrete fluctuators and these fluctuators are not independent. One can associate the fluctuators with certain domains in the NC-CDW phase, and then consider the LFN to be the result of the random processes of the wave de-pinning from various domains. This explains the extremely strong dependence of noise on $V_b$ and $T$. De-pinning of a domain lowers the de-pinning energy of neighboring domains due to their interaction. As a result, the overall activation energy decreases with the increasing number of de-pinned islands and leads to the observed extremely strong noise dependence on $V_b$ and $T$. The size of the fluctuators can be roughly estimated by writing for the fluctuating domain area, $\Delta\Omega$, $\Delta\Omega/\Omega=N\delta I/I$ (where $\Omega$ is the sample area and $N$ is the number of the atomic planes). Assuming that due to the de-pinning event, $I$, changes by $\delta I$, and taking the smallest current step $\delta I\approx$5-10 nA (Fig. 3d), we obtain $(\Delta\Omega)^{1/2}\sim$20 nm – 30 nm. This value is several times larger than the domain sizes determined by scanning tunneling microscopy[13]. Therefore, the smallest observed current steps correspond to a few domains being de-pinned, consistent with the interacting fluctuators picture.

In summary, we demonstrated CDW sliding in an archetypical 2D material via measurements of the low-frequency current fluctuations. We found that electronic noise in 2D CDW systems has unique physical origin associated with evolution and interaction of domains in NC-CDW phase. The LFN spectroscopy provides new insights on transport phenomena in this mixed phase. Considering that many 2D materials undergo transitions to various CDW phases near RT, the obtained results are important not only for understanding the CDW physics but also for future device applications of these materials.

**METHODS**

**1T-TaS$_2$ crystal growth:** The source 1$T$-TaS$_2$ crystals were grown by chemical vapor transport, where the 1$T$ polytype was isolated by fast quenching from the crystal growth temperature. Elemental tantalum (20.4 mmol, Sigma-Aldrich 99.99% purity) and sulfur (41.1 mmol, J.T. Baker >99.9% purity) were ground with mortar/pestle and placed in a 17.8×1.0 cm fused quartz ampule (cleaned overnight with nitric acid followed by 24 h anneal at 900 °C). Elemental iodine (J.T. Baker 99.9% purity) was added (~65 mg for a ~14.0 cm$^3$ ampule volume). The ampule was evacuated and backfilled three times with argon, with cooling to mitigate $I_2$ sublimation. Next the ampule was flame sealed and heated in a two-zone tube furnace at 10 °C min$^{-1}$ to 975 °C (hot zone) and 875 °C (cool zone). These temperatures were held for one week. Then the ampule was removed from the hot furnace and immediately quenched in a water–ice–NaCl bath. The structure and phase purity were verified by powder X-ray diffraction, and the stoichiometry was confirmed with energy dispersive spectroscopy and electron-probe microanalysis (*Supplemental Materials*).

**Device fabrication:** 1*T*-TaS$_2$ thin films were mechanically exfoliated from the bulk crystals and deposited on the Si/SiO$_2$ substrate. To protect the 1*T*-TaS$_2$ thin film from oxidation in air, we used *h*-BN capping immediately after the exfoliation. A thin film of *h*-BN was aligned and transferred on top of 1*T*-TaS$_2$ layer by the dry transfer method. To fabricate 1*T*-TaS$_2$ devices, we used electron beam lithography for defining the electrodes. The channel length was in the range of 1-3 μm, and the width was in the range of 0.8-1.5 μm. We used reactive ion etching to remove part of the *h*-BN capping layer, and deposit metal for forming edge contacts with the 1*T*-TaS$_2$ channel. The electrode materials were 10 nm Ti and 100 nm Au.

**Electronic noise measurements:** The noise spectra were measured with a dynamic signal analyzer (Stanford Research 785) after the signal was amplified by low-noise amplifier (Stanford Research 560). To minimize the 60 Hz noise and its harmonics, we used a battery biasing circuit to apply voltage bias to the devices. The devices were connected with the Lakeshore cryogenic probe station TTPX. All *I-V* characteristics were measured in the cryogenic probe station (Lakeshore TTPX) with a semiconductor analyzer (Agilent B1500). The time domain RTS signals were acquired with the same equipment. In the noise spectra, we removed the data that corresponds to the noise floor of the measurement setup and the 60-Hz electrical grid, if there were any. More details can be found in the *Supplemental Materials*.

**Acknowledgements**

This work was supported, in part, by the National Science Foundation (NSF) Emerging Frontiers of Research Initiative (EFRI) 2-DARE project "Novel Switching Phenomena in Atomic MX$_2$ Heterostructures for Multifunctional Applications" (NSF EFRI-1433395); Semiconductor Research Corporation (SRC) and the Defense Advanced Research Project Agency (DARPA) through the STARnet Center for Function Accelerated nano-Material Engineering (FAME); and the University of California – National Laboratory Collaborative Research and Training Program (UC-NL CRT) project "Mesoscopic Two-Dimensional Materials: From Many-Body Interactions to Device Applications".

**Contributions**

A.A.B. conceived the idea, coordinated the project, and led the experimental data analysis; G.L. designed the experiments, fabricated the devices, conducted noise measurements, and performed experimental data analysis; S.R. performed electronic noise data analysis; T.T.S. supervised material synthesis and contributed to materials characterization; M.A.B. synthesized $1T\text{-}TaS_2$ crystals and performed materials characterization. All authors contributed to the manuscript preparation.

**REFERENCES**

1. Sipos, B. *et al.* From Mott state to superconductivity in 1T-TaS$_2$. *Nat. Mater.* **7,** 960–965 (2008).

2. Sinchenko, A. A., *et al.* Sliding charge-density wave in two-dimensional rare-earth tellurides. *Phys. Rev. B* **85**, 241104 (2012); Sinchenko, A.A *et al.* Dynamical properties of bidirectional charge-density waves in ErTe$_3$. *Phys. Rev. B* **93**, 235141 (2016).

3. Porer, M. *et al.* Non-thermal separation of electronic and structural orders in a persisting charge density wave. *Nat. Mater.* **13,** 857–861 (2014).

4. Stojchevska, L. *et al.* Ultrafast switching to a stable hidden quantum state in an electronic crystal. *Science* **344,** 177–180 (2014).

5. Joe, Y. I. *et al.* Emergence of charge density wave domain walls above the superconducting dome in 1T-TiSe$_2$. *Nat. Phys.* **10,** 421–425 (2014).

6. Campi, G. *et al.* Inhomogeneity of charge-density-wave order and quenched disorder in a high-T$_c$ superconductor. *Nature* **525,** 359–362 (2015).

7. Yu, Y. *et al.* Gate-tunable phase transitions in thin flakes of 1T-TaS$_2$. *Nat. Nanotechnol.* **10,** 270–276 (2015).

8. Feng, Y. *et al.* Itinerant densitywave instabilities at classical and quantum critical points. *Nat. Phys.* **11,** 866–872 (2015).

9. Chatterjee, U. *et al.* Emergence of coherence in the charge-density wave state of 2H-NbSe$_2$. *Nat. Commun.* **6,** 6313 (2015).

10. Samnakay, R. *et al.* Zone-folded phonons and the commensurate–incommensurate charge-density-wave transition in 1T-TaSe2 thin films. *Nano Lett.* **15,** 2965–2973 (2015).

11. Hollander, M. J. *et al.* Electrically driven reversible insulator-metal phase transition in 1T-TaS$_2$. *Nano Lett.* **15,** 1861–1866 (2015).

12. Tsen, A. W. *et al.* Structure and control of charge density waves in two-

dimensional 1T-TaS$_2$. *Proc. Natl. Acad. Sci.* **112,** 15054–15059 (2015).

13.    Ma, L. *et al.* A metallic mosaic phase and the origin of Mott-insulating state in 1T-TaS$_2$. *Nat. Commun.* **7,** 10956 (2016).

14.    Liu, G. *et al.* A charge-density-wave oscillator based on an integrated tantalum disulfide–boron nitride–graphene device operating at room temperature. *Nat. Nanotechnol.* **11,** 845–850 (2016).

15.    Vaskivskyi, I. *et al.* Fast electronic resistance switching involving hidden charge density wave states. *Nat. Commun.* **7,** 11442 (2016).

16.    Vogelgesang, S. *et al.* Phase ordering of charge densitywaves traced by ultrafast low-energy electron diffraction. *Nat. Phys.* (2017). doi:10.1038/NPHYS4309

17.    Grüner, G. The dynamics of charge-density waves. *Rev. Mod. Phys.* **60,** 1129–1181 (1988).

18.    Bloom, I., Marley, A. C. & Weissman, M. B. Nonequilibrium dynamics of discrete fluctuators in charge-density waves in NbSe$_3$. *Phys. Rev. Lett.* **71,** 4385–4388 (1993).

19.    Bloom, I., Marley, A. C. & Weissman, M. B. Discrete fluctuators and broadband noise in the charge-density wave in NbSe3. *Phys. Rev. B* **50,** 5081–5088 (1994).

20.    Grüner, G., Zawadowski, A. & Chaikin, P. M. Nonlinear conductivity and noise due to charge-density-wave depinning in NbSe$_3$. *Phys. Rev. Lett.* **46,** 511–515 (1981).

21.    Bhattacharya, S., Stokes, J. P., Robbins, M. O. & Klemm, R. A. Origin of broadband noise in charge-density-wave conductors. *Phys. Rev. Lett.* **54,** 2453–2456 (1985).

22.    Jamet, J. P., Bouchiat, H., Dumas, J. & Schlenker, C. Broad-band 1/f noise and nonlinear conductivity in an insulating charge-density wave system at low electric-fields. *Europhys. Lett.* **18,** 195–200 (1992).

23.    DiSalvo, F. J. & Fleming, R. M. Search for a sliding charge density wave in

layered compounds. *Solid State Commun.* **35,** 685–687 (1980).

24.     Dutta, P. & Horn, P. M. Low-frequency fluctuation in solids: 1/f noise. *Rev. Mod. Phys.* **53,** 497–516 (1981).

25.     Mitin, V., Reggiani, L. & Varani, L. Generation-recombination noise in semiconductors, 11 – 29, and Lukyanchikova, N.B. Sources of the Lorentzian components in the low-frequency noise spectra of submicron MOSFETs, 201 – 233, in *Noise and Fluctuations Control in Electronic Devices,* ed. Balandin, A. A. (American Scientific Publishers, 2002).

26.     Topalian, Z., Li, S., Niklasson, G. A., Granqvist, C. G. & Kish, L. B. Resistance noise at the metal – insulator transition in thermochromic $VO_2$ films. *J. Appl. Phys.* **117,** 025303 (2016).

27.     Kawasaki, M., Chaudhari, P. & Gupta, A. 1/f noise in $YBa_2Cu_2O_{7-z}$ superconducting bicrystal grain-boundary junctions. *Phys. Rev. Lett.* **68,** 1065–1068 (1992).

28.     Çelik-Butler, Z., Vasina, P. & Amarasinghe, N. V. A method for locating the position of oxide traps responsible for random telegraph signals in submicron MOSFET's. *IEEE Trans. Electron Devices* **47,** 646–648 (2000).

29.     Kirton, M. J. & Uren, M. J. Noise in solid-state microstructures : a new perspective on individual defects, interface states and low-frequency (1/f) noise. *Adv. Phys.* **38,** 376–468 (1989).

30.     Fleetwood, D. M. 1/f noise and defects in microelectronic materials and devices. *IEEE Trans. Electron Devices* **62,** 1462–1486 (2015).

**CAPTIONS**

**Figure 1: Noise spectroscopy as a tool to monitor CDW phenomena in quasi-2D 1$T$-TaS$_2$. a,** $I$-$V$ characteristics of thin-film 1$T$-TaS$_2$ device at RT. The inset shows the device schematic. The abrupt NC-IC and IC-NC CDW phase transitions are seen at voltages $V_H$ (up scan) and $V_L$ (down scan). The red circles indicated the biasing points for LFN measurements. The biasing conditions are divided into four sections I, II, III and IV. The normalized noise spectral densities, $S_I/I^2$, are shown for these regions in the panels c, d, e and f, respectively. **b,** Plot of $S_I/I^2$ at 10 Hz as a function of the bias voltage, $V_b$. Two pronounced local maxima correspond to the de-pinning and NC-IC phase transition. **c,** $S_I/I^2$ for the bias region I. Below $V_b$=150 mV, the noise is of 1/$f$ – type. Above 150 mV, the noise spectrum starts to change to 1/$f^2$ – type, which is indicative of the Lorentzian "tail". **d,** $S_I/I^2$ for the bias region II. At $V_b$=180 mV, the noise spectrum evolves into a Lorentzian shape with a corner frequency $f_c$=10 Hz. As $V_b$ increases, $f_c$ shits to 80.5 kHz at 300 mV. Above $V_b$=300 mV, $f_c$ stops moving to higher frequencies. **e,** $S_I/I^2$ for the bias region III. The noise level reaches its maximum at the NC-IC point. **e,** $S_I/I^2$ for the bias region IV. As $V_b$ drives the 1$T$-TaS$_2$ into the IC-CDW phase, the noise level reduces and its shape returns to 1/$f$ – type. **g,** The corner frequency $f_c$ as function of bias voltage $V_b$.

**Figure 2: Temperature dependent noise in 1$T$-TaS$_2$ across NC-IC CDW phase transition. a,** The resistance of 1$T$-TaS$_2$ device as a function of temperature. The NC-IC phase transition is attained at the same T= 355 K for both biases. The inset shows the domain structure of the NC-CDW phase. **b,** Color map of the frequency normalized noise spectral density $S_I/I^2 \times f$ at $V_b$=30 mV. The evolution of noise from 1/$f$ dependence to Lorentzian at ~320 K is seen as a change of color from blue to red. **c,** The noise level profile at 10 Hz for $V_b$=30 mV. Upon the appearance of Lorentzian noise, $S_I/I^2$ increases by four orders-of-magnitude, and continues to increase towards the NC-IC transition point. Above 355 K, where the NC-IC transition occurs, the noise level sharply decreases. **d,** Color map of $S_I/I^2 \times f$ at $V_b$=60 mV. **e,** The noise profile at 10 Hz for $V_b$=60 mV. The noise level decreases with temperature as the Lorentzian moves to higher frequencies. **f,** Lorentzian corner frequency as function of inverse temperature. For $V_b$=30 mV, $f_c$ is in the range of 43 Hz–64 Hz, and it is independent of temperature. For $V_b$=60 mV, $f_c$ increases with temperature extremely fast. The dashed line shows the fitting used to extract the corresponding activation energy.

**Figure 3: Random telegraph noise and time-domain current fluctuations. a,** The noise spectral density after onset of sliding at different $V_b$. The spectra show "tails" of the Lorentzian noise with $1/f^2$ dependence at $V_b$=100 mV. The corner frequency increases with increasing $V_b$. Inset shows that at some biasing condition, e.g. $V_b$=121 mV, the spectrum can be fitted better with two Lorentzian spectra with different $f_c$, in line with the RTS noise model. **b, c, d,** Time-domain signals at $V_b$ and time scales. Numbers 1, 2, and 3 indicate three characteristic levels contributing to RTN. Note that a small increase of the bias results in a significant change in the noise. The amplitude of the pulses increases and number of fluctuators becomes lager. This is different from classical RTS noise in semiconductor devices.

Figure 1

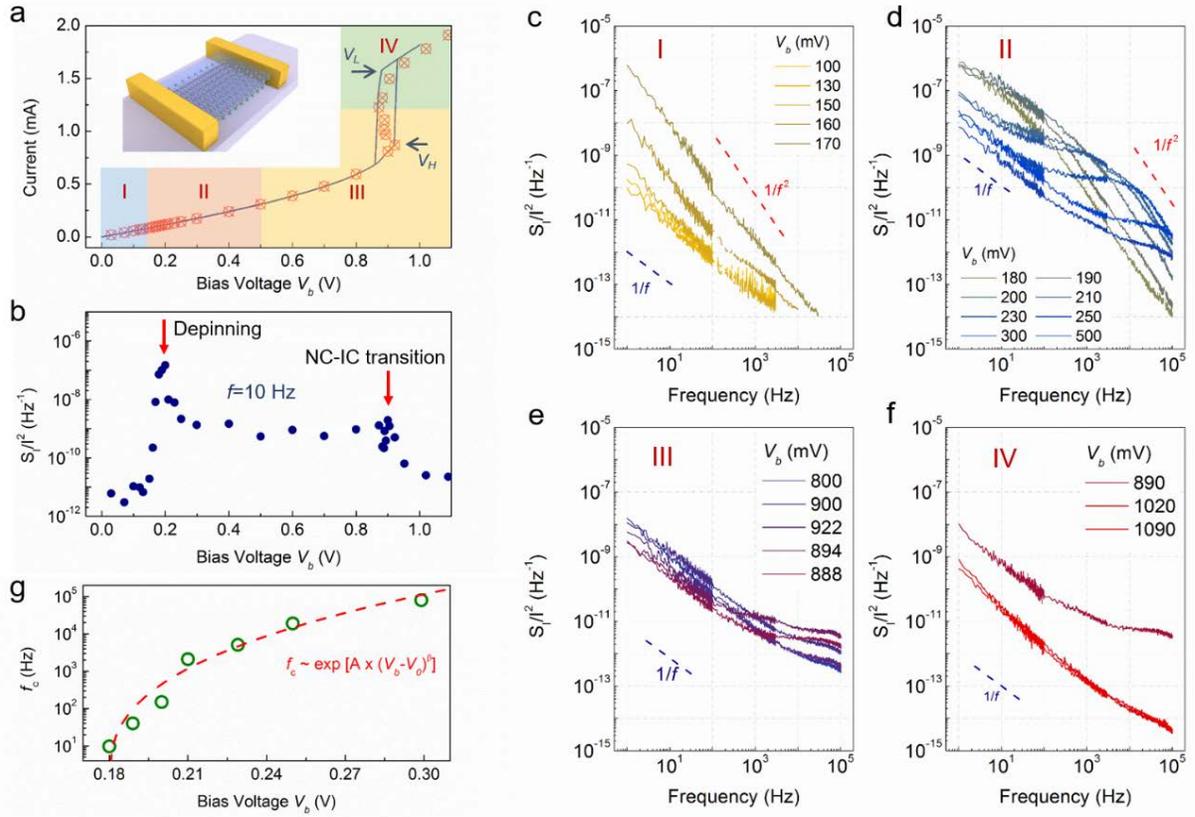

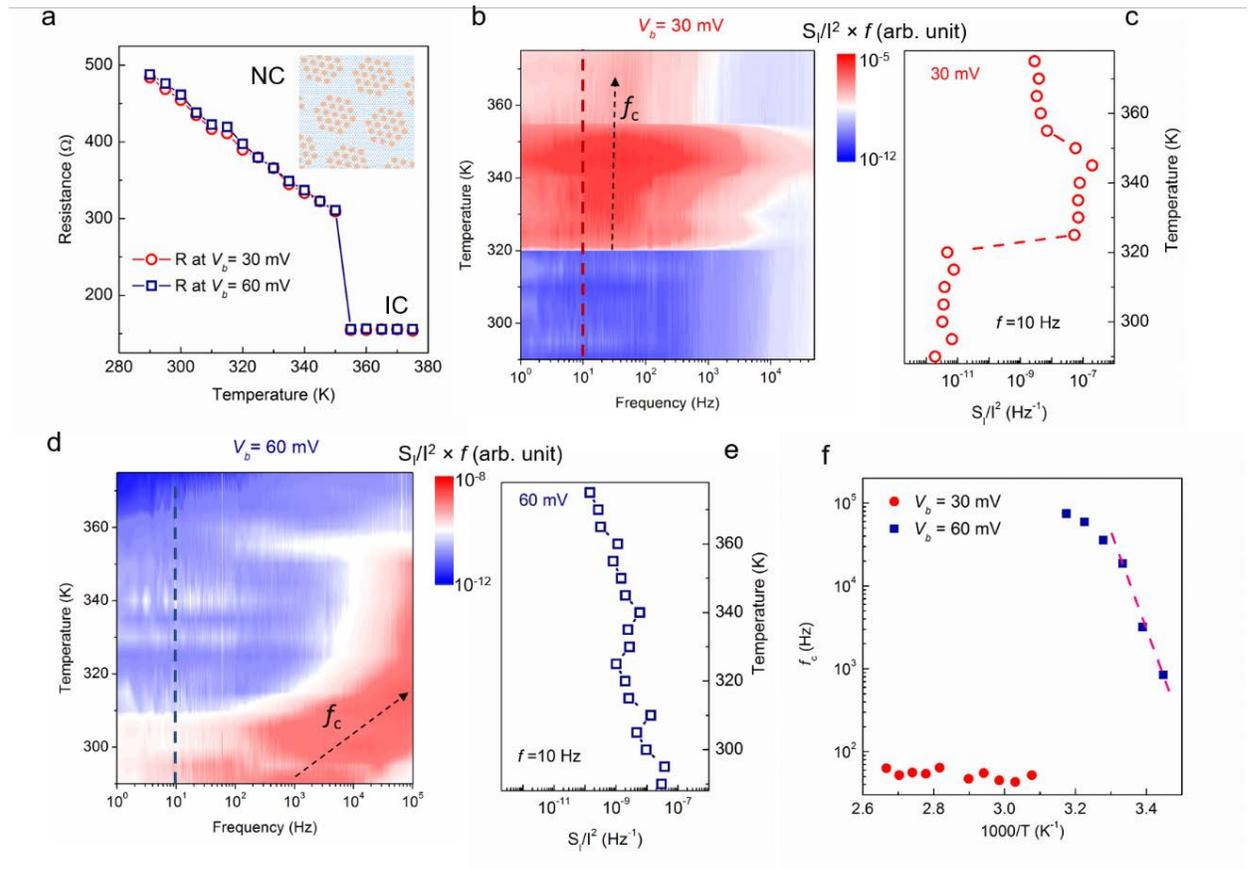

Figure 3

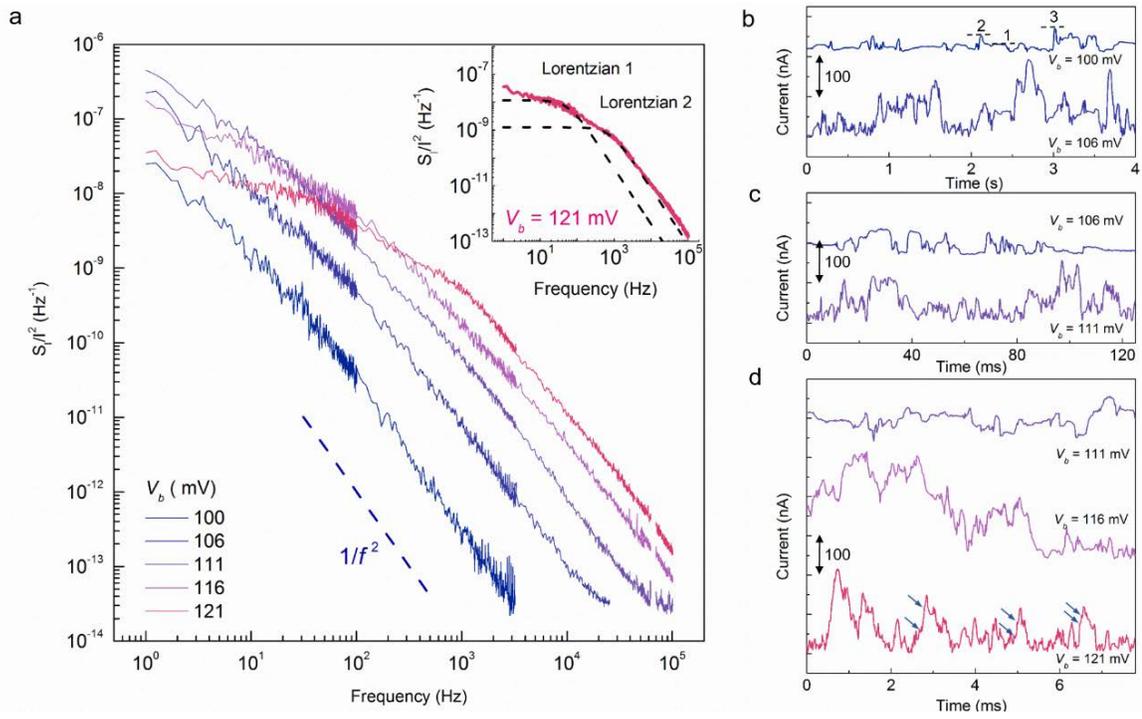